\documentclass{cernrep}
\usepackage{cite,epsfig,amssymb,amsmath}
\newlength{\fdagwidth}
\newlength{\diagupwidth}
\newlength{\stepback}
\newcommand{\fdag}[2][\diagup]{\text{$#2$\settowidth{\fdagwidth}{$#2$}\settowidth{\diagupwidth}{$#1$}\setlength{\stepback}{0.5\fdagwidth}\hspace{-\stepback}\hspace{-0.5\diagupwidth}$#1$\hspace{\stepback}\hspace{-0.5\diagupwidth}}}

\begin{document}
\begin{titlepage} 
\nopagebreak 
{\flushright{ 
        \begin{minipage}{5cm}
       \Large   KA--TP--05--2006  \\            
        {\tt hep-ph/0605117}\hfill \\ 
        \end{minipage}        } 
 
} 
\vfill 
\begin{center} 
{\LARGE \bf \sc 
 \baselineskip 0.9cm 
Higgs + 2 Jets as a Probe for CP
  Properties~\footnote{Contribution to the Proceedings of the Workshop
                 on CP Violation and Non-standard Higgs Physics, CERN}
} 
\vskip 0.5cm  
{\large   
V.~Hankele, G.~Kl\"amke, and D.~Zeppenfeld
}   
\vskip .2cm  
{\it Institut f\"ur Theoretische Physik, 
        Universit\"at Karlsruhe, P.O.Box 6980, 76128 Karlsruhe, Germany}

 \vskip 
1.3cm     
\end{center} 
 
\nopagebreak 
%\vfill 
%\vskip 3cm 
\begin{abstract}
 Azimuthal angle correlations of the jets in Hjj events at the LHC provide a
 probe of the CP nature of Higgs couplings to gauge bosons. In weak boson
 fusion the HWW and HZZ couplings are tested. Gluon fusion processes probe
 the tensor structure of the effective Hgg vertex and thus the CP
 nature of the dominant quark couplings.
\end{abstract} 
\vfill 
%\today \timestamp \hfill 
\vfill 

% PACS: 14.80.Bn
\end{titlepage} 
% 
%\newpage               
%
%%%%%%%%%%%%%%%%%%%%%%%%%%%%%%%%%%%%%%%%%%%%%%%%%%%%%%%%%%%%%%%%%
%           

%---------------------------------------------------------------------

%% \title{Higgs + 2 Jets as a Probe for CP
%%   Properties%~\footnote{Contribution to the Proceedings of the Workshop
%%             %    on CP Violation and Non-standard Higgs Physics, CERN}
%% }

%% \author{V.~Hankele, G.~Kl\"amke, and D.~Zeppenfeld}
%% \institute{Institut f\"ur Theoretische Physik, 
%%         Universit\"at Karlsruhe, P.O.Box 6980, 76128 Karlsruhe, Germany}

%% \maketitle

%% \begin{abstract}
%%  Azimuthal angle correlations of the jets in Hjj events provide a 
%%  probe of the CP nature of Higgs couplings to gauge bosons. In weak boson 
%%  fusion the HWW and HZZ couplings are tested. Gluon fusion processes probe 
%%  the tensor structure of the effective Hgg vertex and thus the CP 
%%  nature of the dominant quark couplings.
%% \end{abstract}

%Contact: dieter@particle.uni-karlsruhe.de\\

\noindent
At the LHC, one would like to experimentally determine the CP nature of
any previously discovered (pseudo)scalar resonance. Such measurements
require a complex event structure in order to provide the
distributions and correlations which can distinguish between CP-even and
CP-odd couplings. This can either be done by considering
decays, e.g. $H\to ZZ\to  l^+l^- l^+l^-$ and the correlations of the
decay leptons~\cite{Choi:2002jk,Buszello:2002uu}, 
or one can study correlations arising in the production 
process.  Here the azimuthal angle correlations between the two
additional jets in $Hjj$ events have emerged as a promising
tool~\cite{Plehn:2001nj}. In the following we consider the prospects for
using $\Phi jj$ events at the LHC, where $\Phi$ stands for either a CP
even boson, H, or a CP odd state, A. Two production processes are considered.
The first is vector boson fusion (VBF), i.e. the electroweak process $qQ\to
qQ\Phi$ (and crossing related ones) where $\Phi$ is radiated off a
$t$-channel electroweak boson. The second is gluon fusion where $\Phi$ is
produced in QCD dijet events, via the insertion of a heavy quark loop
which mediates $gg\to\Phi+0,1,2$ gluons. 
 
The CP properties of a scalar field are defined by its couplings and
here we consider interactions with fermions as well as gauge
bosons. Within renormalizable models the former are given
by the Yukawa couplings
%---------- YR only
%of Eq.~(\ref{}). In our numerical analysis we
%consider top quark couplings of SM strength, i.e. $\chi_t = g_t/g_t^{SM} =1$
%for a scalar field, $H$, and  $\chi_t = i$ for a pseudoscalar, $A$.
%---------  end YR
\begin{equation}
{\cal L}_Y = {y}_{f}\bar{\psi}H\psi + 
\tilde{y}_{f}\bar{\psi}Ai\gamma_5\psi \;,
\label{eq:Yukawa}
\end{equation}
where $H$ (and $A$) denote (pseudo)scalar fields which couple to
fermions $\psi=t,\;b, \;\tau$ etc. In our numerical analysis we
consider couplings of SM strength, $y_f=\tilde y_f = m_f/v = y_{SM}$. 
Via these Yukawa couplings, quark loops induce
effective couplings of the (pseudo)scalar to gluons which, for
(pseudo)scalar masses well below quark pair production threshold, can be
described by the effective Lagrangian
%--------- YR only
%\begin{equation}
%{\cal L}_{\rm eff} = {\rm Re}\chi_t \cdot 
%\frac{\alpha_s}{12\pi v} \cdot H \, G_{\mu\nu}^a\,G^{a\,\mu\nu} + 
%{\rm Im}\chi_t \cdot
%\frac{\alpha_s}{16\pi v} \cdot A \, 
%G^{a}_{\mu\nu}\,G^{a}_{\rho\sigma}\varepsilon^{\mu\nu\rho\sigma}\;.
%\label{eq:ggS}
%\end{equation}
%-------- end YR only
\begin{equation}
{\cal L}_{\rm eff} = \frac{y_f}{y_{SM}}\cdot 
\frac{\alpha_s}{12\pi v} \cdot H \, G_{\mu\nu}^a\,G^{a\,\mu\nu} + 
\frac{\tilde y_f}{y_{SM}} \cdot
\frac{\alpha_s}{16\pi v} \cdot A \, 
G^{a}_{\mu\nu}\,G^{a}_{\rho\sigma}\varepsilon^{\mu\nu\rho\sigma}\;.
\label{eq:ggS}
\end{equation}
Similar to the $\Phi gg$ coupling, Higgs couplings to $W$ and $Z$ bosons
will also receive contributions from heavy particle loops which can be
parameterized by the effective Lagrangian
\begin{equation}
{\cal L}_5 =
  \frac{f_{\rm e}}{\Lambda_{5}} \; H \;
    \vec{W}_{\mu\nu} \vec{W}^{\mu\nu}
+ \frac{f_{\rm o}}{\Lambda_{5}} \; A \;
   \vec{W}_{\mu\nu}
   \vec{W}_{\rho\sigma}\,\frac{1}{2}\varepsilon^{\mu\nu\rho\sigma}  \;.
\label{eq:d5}
\end{equation}

For most models, one expects a coupling strength of order $f_i/\Lambda_5 \sim
\alpha/(4\pi v)$ for these dimension 5 couplings and, hence, cross
section contributions to vector boson fusion processes which are
suppressed by factors $\alpha/\pi$ (for interference effects with SM
contributions) or $(\alpha/\pi)^2$ compared to those mediated by the
tree level $HVV\; (V=W,\;Z)$ couplings of the SM. However, together with
the tree level 
couplings, the effective Lagrangian of Eq.~(\ref{eq:d5}) has the virtue
that it parameterizes the most general $\Phi VV$ coupling which can
contribute in the vector boson fusion process $qQ\to qQ\Phi$ and, thus,
it is a convenient tool for phenomenological discussions and for
quantifying, to what extent certain couplings can be excluded experimentally.
Neglecting terms which vanish upon contraction with the conserved quark
currents, the most general tensor structure
for the fusion vertex $V^\mu(q_1)V^\nu(q_2)\to \Phi$ is given by
\begin{equation}
T^{\mu\nu}(q_1,q_2) = a_1(q_1,q_2)\;g^{\mu\nu} +
a_2(q_1,q_2)\;\left[ q_1\cdot q_2 g^{\mu\nu} - q_2^\mu q_1^\nu \right] +
a_3(q_1,q_2)\;\ \varepsilon^{\mu\nu\alpha\beta} q_{1\alpha}q_{2\beta} \;.
\end{equation}
Here the $a_i(q_1,q_2)$ are scalar form factors, which, in the low
energy limit, are given by the effective
Lagrangian of Eq.~(\ref{eq:d5}). One obtains, e.g. for the $W^+W^-\Phi$
coupling, $a_2 = -2f_e/\Lambda_5$ and $a_3 = 2f_o/\Lambda_5$, while 
$a_1 = 2m_W^2/v$ is the SM vertex.  

\begin{figure}[htb]
\centerline{
\includegraphics[height=7.2cm, width=7.8cm]{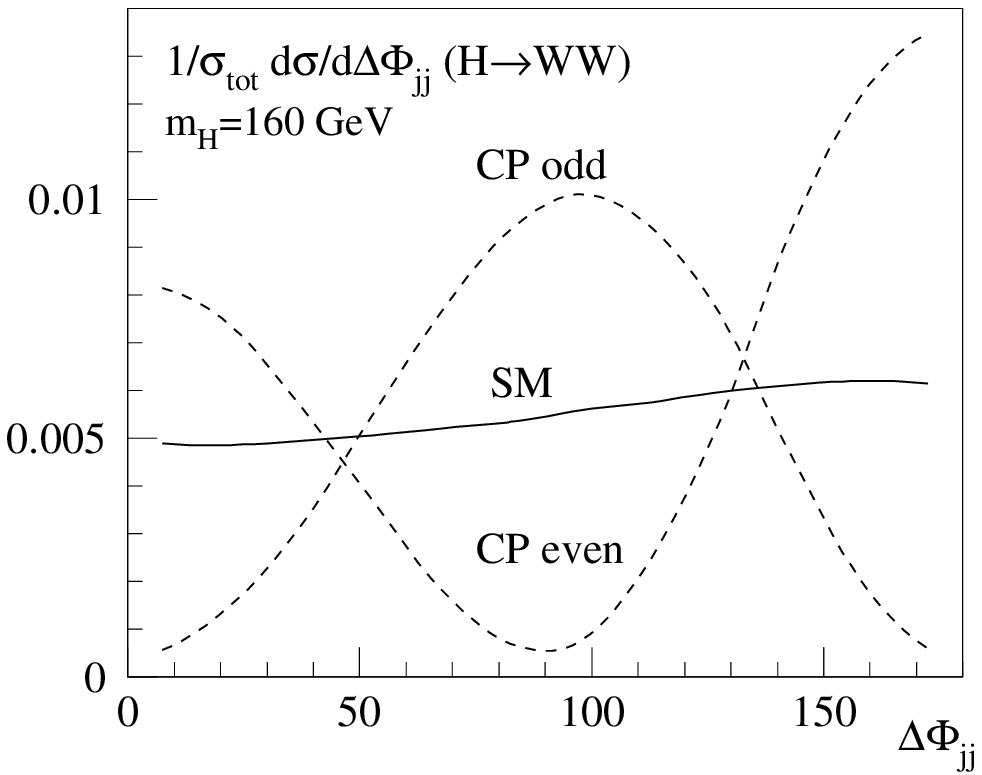} \hspace*{0.3cm}
\includegraphics[scale=0.4]{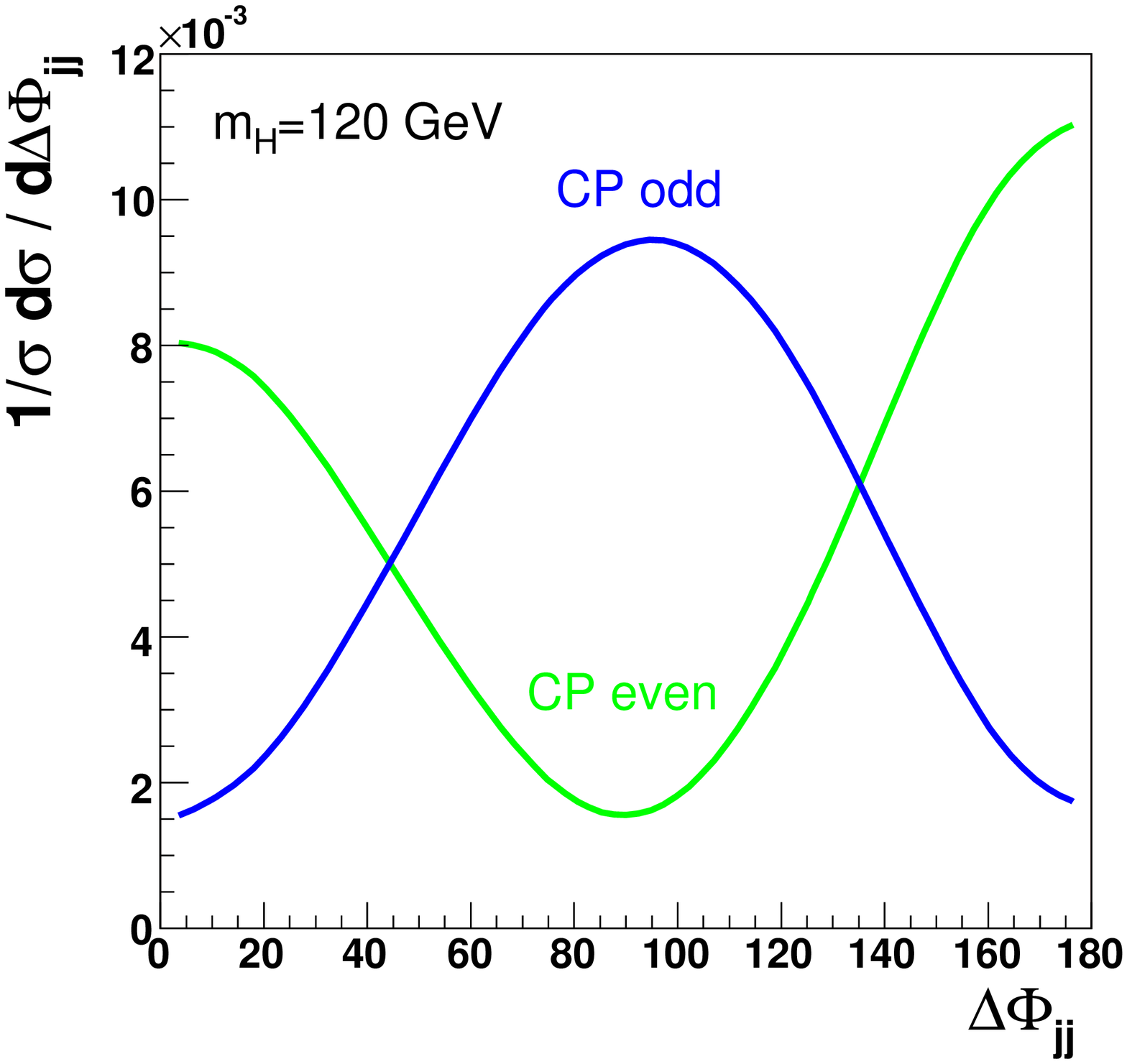}
}
\caption[]{\label{fig:phijj} {\it Left:} Normalized distributions of the
  azimuthal angle between the two 
  tagging jets, for the $\Phi\to WW\to e\mu\fdag{p}_T$ signal in vector boson
  fusion at $m_\Phi = 160$~GeV, from Ref.~\protect\cite{Plehn:2001nj}. 
Curves are for the SM and for single D5 operators as given in
  Eq.~(\ref{eq:d5}), after cuts as in
  Ref.~\protect\cite{Rainwater:1999sd}.
{\it Right:} The same for Higgs production in gluon fusion at
  $m_\Phi=120$~GeV. Curves are for CP-even and CP-odd $\Phi tt$ coupling.
}

\end{figure}

The CP-even and CP-odd couplings of Eqs.~(\ref{eq:ggS},\ref{eq:d5}) lead
to characteristic azimuthal angle correlations of the two jets in $\Phi jj$
production processes. Normalized distributions of the azimuthal angle
between the two jets, $\triangle \phi_{jj}$, are shown in
Fig.~\ref{fig:phijj} for vector boson fusion (left panel) and for gluon
fusion processes (right panel) leading to $\Phi jj$ events: A CP-odd
coupling suppresses the cross section for planar events because
the epsilon tensor contracted with the four linearly dependent momentum
vectors of the incoming and outgoing partons disappears. For a CP-even
coupling the dip, instead, appears at 90
degrees~\cite{Plehn:2001nj,DelDuca:2001fn}. Unfortunately, when  
both CP-even and CP-odd couplings are present simultaneously, the two 
$\triangle \phi_{jj}$ distributions simply add, i.e. one does not 
observe interference effects. The dip-structure which is present for pure
couplings is, thus, washed out.

\begin{figure}[bht]

\centerline{
\includegraphics[height=7.2cm, width=7.5cm]{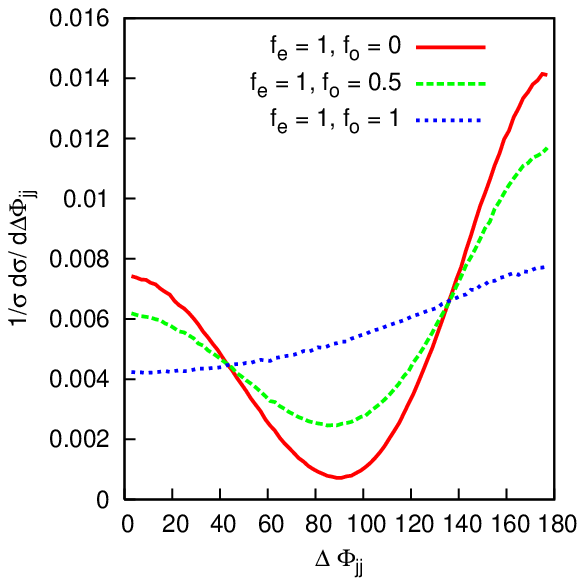} \hspace*{0.3cm}
\includegraphics[height=7.2cm, width=7.8cm]{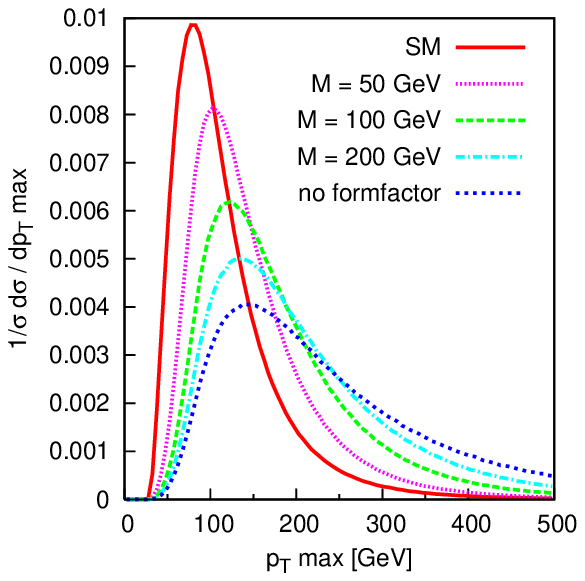}
}

\caption[]{\label{fig:wbf} Normalized distributions of the tagging jets
  in EW vector boson fusion with anomalous couplings and for a Higgs
  mass of $m_\Phi =120$~GeV. Typical VBF cuts of $ p_{Tj} > 30\, {\rm GeV},\
  |\eta_j| < 4.5,\ |\eta_{j_1}-\eta_{j_2}| > 4.0,\ m_{jj} > 600~{\rm GeV}$ 
  are applied.  
{\it Left:}   Azimuthal angle distribution between the
  two tagging jets, for different strengths of the operators of
  Eq.~(\ref{eq:d5}).
{\it Right:}  Transverse momentum distribution of the
  hardest tagging jet for $f_e = f_o = 1$ and a form factor as in
  Eq.~(\ref{eqn:C0}). The ``no formfactor'' curve corresponds to the
  limit $M\to\infty$, i.e. a constant $a_i$.  
}
\end{figure}

This behavior is demonstrated in Fig.~\ref{fig:wbf}. For CP-even
  and CP-odd couplings of the same strength, i.e. $f_e = f_o$, the
  azimuthal angle distribution is very similar to the SM case. However,
  in order to test the presence of anomalous couplings in such cases,
  other jet distributions can be used, e.g. transverse
  momentum distributions. The $\triangle \phi_{jj}$ distribution
  is quite insensitive to variations of form factors, NLO corrections
  and the like~\cite{Figy:2004pt}. On the other hand, $p_T$
  distributions depend strongly on form factor effects. We study these
  effects for a particular parameterization of the momentum dependence:
\begin{equation}
a_2(q_1,q_2) = a_3(q_1,q_2)  \sim M^2 \; C_0 \ ( \ q_1, q_2, M \ )\;,
\label{eqn:C0}
\end{equation}
where $C_0$ is the familiar Passarino-Veltman scalar three-point 
function~\cite{Passarino:1978jh}.
This ansatz is motivated by the fact that the $C_0$ function naturally
appears in the calculation of one-loop triangle diagrams, where the mass
scale M is given by the mass of the heavy particle in the loop. 
As can be seen in the right panel of Fig.~\ref{fig:wbf}, even for a mass
scale M of the order of 50 GeV the anomalous couplings produce a harder
$p_T$ distribution of the tagging jets than is expected for SM
couplings. Thus it is possible to experimentally distinguish EW vector
boson fusion as predicted in the SM from loop induced $WW\Phi$ or
$ZZ\Phi$ couplings by the shape analysis of distributions alone.

Let us now consider the gluon fusion processes where, for $\Phi tt$
couplings of SM strength, one does expect observable event rates from the
loop induced effective $\Phi gg$ couplings~\cite{DelDuca:2001fn}.
In order to assess the visibility of the CP-even vs. CP-odd signatures
of the azimuthal jet correlations at the LHC, we consider 
Higgs + 2 jet production with 
the Higgs decaying into a pair of $W$-bosons which further decay
leptonically, $\Phi \rightarrow W^+W^-\rightarrow \ell^+\ell^-\nu\bar{\nu}$. We
only consider electrons and muons ($\ell=e^\pm,\,\mu^\pm$) in the final
state. The Higgs-mass is set to $m_\Phi = 160$~GeV. 
From previous studies on Higgs production in
vector boson fusion~\cite{Rainwater:1999sd} the main backgrounds are
known to be top-pair production i.e. $pp\rightarrow
t\bar{t},\,t\bar{t}j,\,t\bar{t}jj$~\cite{Kauer:2002sn}. The three cases
distinguish the number of $b$ quarks which emerge as tagging jets. The
$t\bar{t}$ case corresponds to both bottom-quarks from the 
top-decays being identified as forward tagging jets, for $t\bar{t}j$
production only one tagging jet arises from a $b$ quark, while the $t\bar{t}jj$
cross section corresponds to both tagging jets arising from massless
partons. Further backgrounds arise from QCD induced $W^+W^-$ + 2
jet production and electroweak $W^+W^-jj$ production. These backgrounds are
calculated as in Refs.~\cite{Barger:1989yd} and~\cite{Jager:2006zc},
respectively. In the EW $W^+W^-jj$ background, 
Higgs production in VBF is included, i.e. the VBF Higgs signal is
considered as a background to 
the observation of $\Phi jj$ production in gluon fusion. We do not
consider backgrounds from $Zjj,\; Z\to\tau\tau$ and from $b\bar bjj$
production because they have been shown to be small in the
analyses of Refs.~\cite{Rainwater:1999sd,Kauer:2000hi}.

\begin{table}[htb]
  \caption{Signal rates and background cross sections for $m_\Phi=160\,{\rm
      GeV}$. Results are given for the inclusive cuts of Eq.~(\ref{eqn:incl}),
    the additional selection cuts of Eq.~(\ref{eqn:sel}) and b-quark
    identification as discussed in the text, and with the additional 
    $\Delta\eta_{jj}$ cut of Eq.~(\ref{eqn:eta}) which improves the 
    sensitivity to the CP nature of the $\Phi tt$ coupling. The events 
    columns give the expected number of events for 
    ${ \cal L}_{int}=30\,{\rm fb}^{-1}$. }
\begin{center}
\begin{tabular}{|c|c|c|c|c|c|}
 \cline{2-6}
\multicolumn{1}{c|}{} & inclusive cuts & \multicolumn{2}{c|}{selection
  cuts} & \multicolumn{2}{c|}{selection
  cuts + Eq.~(\ref{eqn:eta}) }\\
\hline
process & $\sigma$ [fb] & $\sigma$ [fb] & events / {$30\,{\rm fb}^{-1}$}
&  $\sigma$ [fb] & events / {$30\,{\rm fb}^{-1}$}\\
\hline\hline
GF $pp \rightarrow \Phi +j j$ & 121.2 & {39.2} &{1176} & 13.1 & 393\\
VBF $pp \rightarrow W^+W^- +j j$ & 75.2 & {20.8} & {624} & 17.4 & 521\\
$pp \rightarrow t \bar{t}$ & 6832 & {29.6} & {888} & 2.0 & 60 \\
$pp \rightarrow t \bar{t}+ j$ & 9712 & {56.4} & {1692} & 15.6 & 468\\
$pp \rightarrow t \bar{t}+ jj $ & 1200 & {8.8} & {264} & 3.2 & 97\\
QCD $pp \rightarrow W^+W^- +j j$ & 364 & {15.2} & {456} & 3.9 & 116\\
\hline
\end{tabular}
\end{center}
\label{tab:xs}
\end{table}

\begin{figure}[htb]
\centerline{
\includegraphics[scale=0.4]{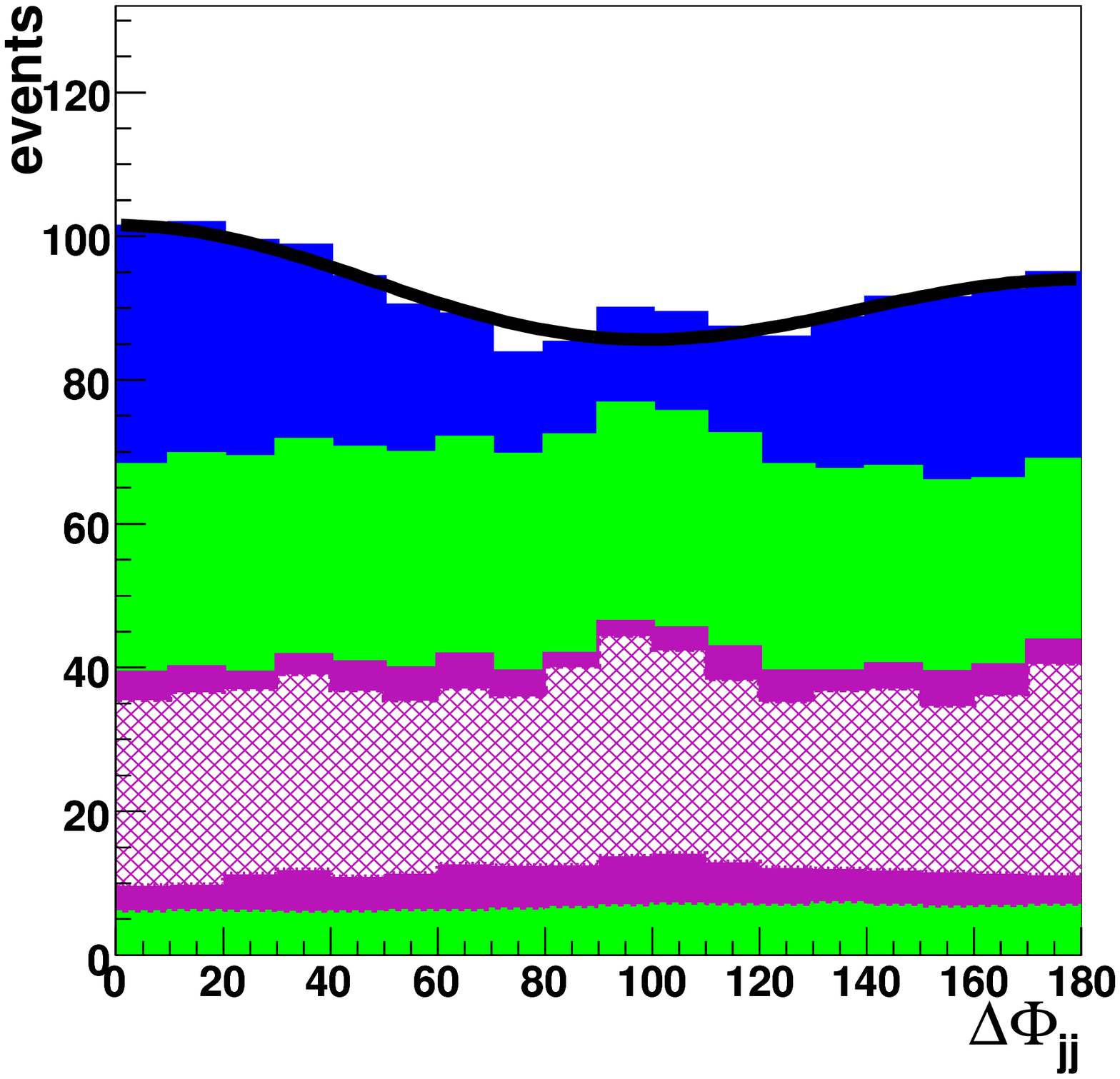} 
\includegraphics[scale=0.4]{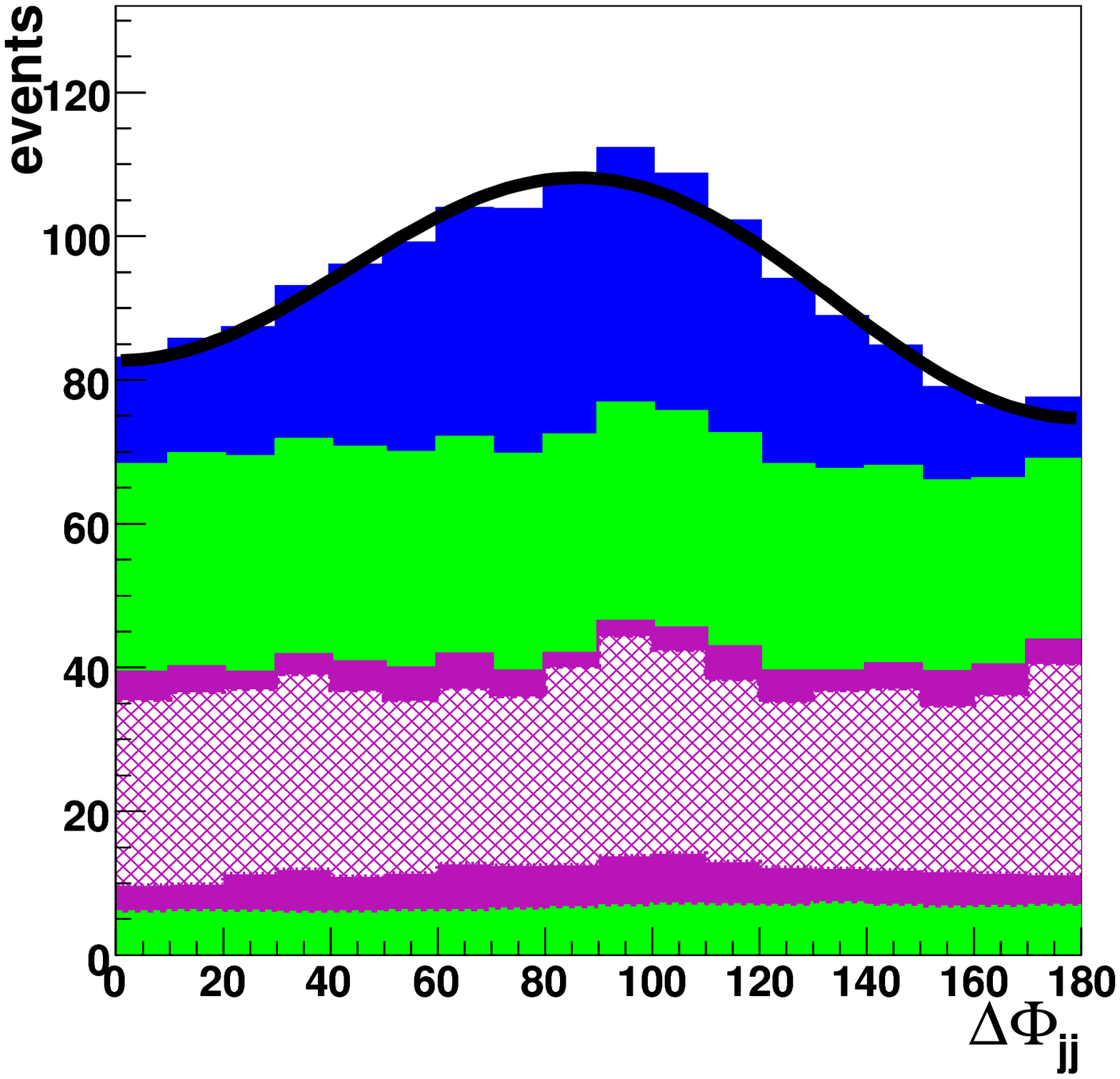}
}
\caption[]{\label{fig:events} Distribution of the azimuthal angle
  between the tagging jets  in $\Phi jj$ events
  for a CP-even $\Phi tt$ coupling ({\it left}) and a CP-odd coupling
  ({\it right}). Shown are expected signal and background events per 
  10 degree bin for 
  $\Phi \rightarrow W^+W^-\rightarrow \ell^+\ell^-\nu\bar{\nu}$ and 
  ${\cal L}_{int}=30\,{\rm fb}^{-1}$ for
  the cuts of Eqs. (\ref{eqn:incl}, \ref{eqn:sel}, \ref{eqn:eta}) and an applied b-veto.
  Processes from top to bottom: gluon fusion (signal), VBF, $t\bar{t}$,
  $t\bar{t}j$, $t\bar{t}jj$, QCD $WWjj$. $m_{\Phi}=160$~GeV is assumed.
}
\end{figure}

The inclusive cuts in Eq.~(\ref{eqn:incl}) reflect the requirement
that the two tagging jets and two charged leptons are observed inside
the detector, and are well-separated from each other.
\begin{equation*}
p_{Tj} > 30\, {\rm GeV},\qquad|\eta_j| < 4.5,\qquad |\eta_{j_1}-\eta_{j_2}| > 1.0
\end{equation*}
\begin{equation}
p_{T\ell} > 10\, {\rm GeV},\qquad|\eta_\ell| < 2.5,\qquad \Delta R_{j\ell} > 0.7
\label{eqn:incl}
\end{equation}
The resulting cross sections for these cuts are shown in Table
\ref{tab:xs}. The signal cross section of 121~fb (which includes the
branching ratios into leptons) is quite sizeable. 
The QCD $WWjj$ cross section is about 3 times higher whereas the VBF
process reaches $2/3$ of the signal rate. The worst source of background
arises from the $t\bar{t}$ processes, with a total cross section of more than
17~pb.\\ 
In order to improve the signal to background ratio the following
selection cuts are applied:
\begin{equation*}
p_{T\ell} > 30\, {\rm GeV},\qquad m_{\ell\ell} < 75\, {\rm GeV},\qquad
\Delta R_{\ell\ell} < 1.1
\end{equation*}
\begin{equation}
m_{T}^{WW} < 170\, {\rm GeV},\qquad m_{\ell\ell} < 0.5\cdot m_{T}^{WW}\;.
\label{eqn:sel}
\end{equation}
Here, the transverse mass of the dilepton-$\fdag{\vec{p}}_T$ system is
defined as~\cite{Rainwater:1999sd}
\begin{equation}
m_T^{WW} = \sqrt{(\fdag{E}_T+E_{T,{\ell\ell}})^2-
({\vec{p}}_{T,{\ell\ell}}+\fdag{\vec{p}}_T)^2}
\end{equation}
in terms of the invariant mass of the two charged lepton and the 
transverse energies 
\begin{equation}
E_{T,\ell\ell} = (p_{T,\ell\ell}^2 + m_{\ell\ell}^2)^{1/2},\qquad
\fdag{E}_T = (\fdag{p}_T^2 + m_{\ell\ell}^2)^{1/2} .
\end{equation}
%{\it ...Explanation of the cuts...}\\
In addition to these cuts we make use of a b-veto to reduce the
large top-background. We reject all events where at least one jet is identified
as a b-jet. Using numbers from Ref.~\cite{Weiser:2006md}, we assume b-tagging
efficiencies in the range of $60\%-75\%$ (depending on b-rapidity and
transverse momentum) and an overall mistagging probability of $10\%$ for
light partons.\\
With the selection cuts (\ref{eqn:sel}) and the b-veto the backgrounds
can be strongly suppressed. Table \ref{tab:xs} shows the resulting
cross sections and the expected number of events for an integrated
luminosity of ${\cal L}_{int}=30\,{\rm fb}^{-1}$. The signal rate is reduced
by a factor of 3 but the backgrounds now have cross sections of the same
order as the signal. The largest background still comes from the
$t\bar{t}$ processes,
especially $t\bar{t}+1j$. For $30\,{\rm fb}^{-1}$ we get about $1000$
signal events on top of $4000$ background events. This corresponds to a
purely statistical significance of the gluon fusion signal of 
$S/\sqrt{B}\approx 18$ and a sufficient number of events 
to analyze the azimuthal jet correlations.

Fig.~\ref{fig:events} shows the expected $\triangle \phi_{jj}$
distribution for $30\,{\rm fb}^{-1}$. Plotted are
signal events on top of the various backgrounds. 
An additional cut on the rapidity gap between the jets 
\begin{equation}
|\eta_{j_1}-\eta_{j_2}| > 3.0
\label{eqn:eta}
\end{equation}
has been applied. It enhances the shape of the distribution that is
sensitive to the nature of the $\Phi tt$ coupling. Clearly visible, the
distribution for the CP-even coupling has a slight minimum at $\triangle
\phi_{jj} = 90^\circ$ whereas for the CP-odd case there is a pronounced
maximum. In order to quantify this, we define the fit-function
\begin{equation}
f(\triangle \phi) = C\cdot(1+A\cdot\cos{2\triangle
  \phi}+B\cdot\cos{\triangle \phi})
\end{equation}
with free parameters $A$, $B$, $C$.
The fit is shown as black curves in Fig. \ref{fig:events}. The parameter
$A$ is now a measure for the $\triangle \phi_{jj}$ asymmetry,
i.e. whether there is a CP-even or CP-odd $\Phi tt$ coupling. The fitted
values are $A = 0.064 \pm 0.035$ for the CP-even and $A=-0.157 \pm
0.034$  for the CP-odd case, while $A_{B} = -0.039 \pm 0.040$ for the 
sum of all backgrounds. Defining a significance $s$ as
\begin{equation}
s = \frac{(A_{S+B}-A_{B})}{\Delta A_{S+B}}\; ,
\end{equation}
we get $s = 3.0$ and $s = -3.4$ for the CP-even and CP-odd case,
respectively.
Thus, a distinction of a CP-odd and CP-even $\Phi tt$ coupling is possible
at a $6\sigma$ level for the considered process and a Higgs mass of
160~GeV. This implies that, at least for favorable values of the Higgs
boson mass, (i) an effective separation of VBF and gluon fusion sources
of $\Phi jj$ events is possible and (ii) the CP nature of the $\Phi tt$
coupling of Eq.~(\ref{eq:Yukawa}) can be determined at the LHC.

%---------------------------------------------------------------------
\section*{Acknowledgments}
%---------------------------------------------------------------------

This research was supported by the Deutsche
Forschungsgemeinschaft in the Sonderforschungsbereich/Transregio 
SFB/TR-9 ``Computational Particle Physics'' and in the Graduiertenkolleg
"High Energy Physics and Particle Astrophysics".

\bibliography{hxjj}

\end{document}